\documentclass[pra,onecolumn,12pt]{revtex4}
\usepackage{amsfonts}
\usepackage{amsmath}
\usepackage{amssymb}
\usepackage{graphicx}

\input{tcilatex}
\begin{document}

\title{Generation of decoherence-free displaced squeezed states of radiation
fields and a squeezed reservoir for atoms in cavity QED }
\author{T. Werlang$^{1}$, R. Guzm\'{a}n$^{2}$, F. O. Prado$^{1}$, and C. J.
Villas-B\^{o}as$^{1}$}
\affiliation{$^{1}$Departamento de F\'{\i}sica, Universidade Federal de S\~{a}o Carlos,
13595-905, S\~{a}o Carlos, Brasil}
\affiliation{$^{2}$Departamento de Ciencias F\'{\i}sicas, Universidad de La Frontera,
Casilla Postal 54-D, Temuco, Chile}

\begin{abstract}
We present a way to engineer an effective anti-Jaynes-Cumming and a
Jaynes-Cumming interaction between an atomic system and a single cavity mode
and show how to employ it in reservoir engineering processes. To construct
the effective Hamiltonian, we analyse considered the interaction of an
atomic system in a $\Lambda $ configuration, driven by classical fields,
with a single cavity mode. With this interaction, we firstly show how to
generate a decoherence-free displaced squeezed state for the cavity field.
In our scheme, an atomic beam works as a reservoir for the radiation field
trapped inside the cavity, as employed recently by S. Pielawa et al. [Phys.
Rev. Lett. \textbf{98}, 240401 (2007)] to generate an
Einstein-Podolsky-Rosen entangled radiation state in high-Q resonators. In
our scheme, all the atoms have to be prepared in the ground state and, as in
the cited article, neither atomic detection nor precise interaction times
between the atoms and the cavity mode are required. From this same
interaction, we can also generate an ideal squeezed reservoir for atomic
systems. For this purpose we have to assume, besides the engineered
atom-field interaction, a strong decay of the cavity field (i.e., the cavity
decay must be much stronger than the effective atom-field coupling). With
this scheme, some interesting effects in the dynamics of an atom in a
squeezed reservoir could be tested.
\end{abstract}

\maketitle

\section{Introduction}

The impressive experimental progress in manipulation of the interaction
between light and atoms has led to a better understanding of several
fundamentals of quantum theory and also the development of the area known as
quantum information theory \cite{nielsen}. The experimental verification of
the granular nature of the radiation field \cite{granular-light} or of the
motion of a single trapped ion \cite{granular-ion}, the study of the
decoherence process of a Schr\"{o}dinger cat superposition state \cite%
{decoherence-haroche} and the violation of the Bell inequalities \cite%
{bell-experiment}, which reveals the non-local character of quantum
phenomena, are some examples of fundamentals of physics recently
investigated through the manipulation of radiation field states by atoms or
vice-versa. On the other hand, the implementation of quantum logic gates in
trapped ions \cite{gates-ions} or in cavity quantum electrodynamics (QED) 
\cite{gates-cavities} and atomic teleportation \cite{teleportation} have
contributed to the rapid development of the quantum information area \cite%
{nielsen}. Through the precise manipulation of the atom-field interaction,
many quantum states of light such as the Schr\"{o}dinger cat states \cite%
{haroche-RMP2001} and Fock states \cite{fock-state-cavity} have been
generated in cavity QED. However, some non-classical states, such as the
squeezed states \cite{review-dodonov} and the two-mode squeezed state \cite%
{two-mode-squeezed}, have not been attained experimentally so far, either in
the microwave or in the optical regimes in cavity QED, through the
interaction of atoms and cavity modes. The experimental generation of these
states is of great interest since they could be used to teste the
fundamentals of theoretical physics and to achieve quantum communication.
For instance, the single-mode squeezed states could be used to verify the
sub-Poissonian statistics of the radiation field \cite{poissonian}, to
measure gravitational waves \cite{gravitational-waves}, and for optical
communication, through improvement of the signal-to-noise ratio \cite%
{squeezed-communication}. The generation of multimode squeezed light would
also be useful for manipulating the dynamics of two-level atoms. As pointed
out in \cite{gardiner-PRL1986, scully-livro}, the interaction of a two-level
atom with multimode squeezed light, which works as a squeezed reservoir, can
produce some interesting effects in atomic dynamics such as suppression
(enhancement) of decay of the in-phase (out-of-phase) components of atomic
polarization and a line narrowing in resonance fluorescence and absorption
spectra. Therefore, the generation of robust squeezed states of the
radiation field and the generation of a squeezed reservoir for two-level
atoms may help us to deepen our understanding of the quantum nature of the
light, the properties of atoms and the atom-field interaction itself.

Concerning radiation squeezed states, we find some theoretical schemes in
the literature for the generation of these states in the cavity QED context
employing the interaction of three-level atoms \cite{three-level-squeezing}
or even a single two-level atom \cite{prado-pra2006} with a trapped field
inside a high-finesse cavity. There are some theoretical proposals,
employing the manipulation of the interaction between a three-level atom in
a $\Lambda $ configuration and a single cavity mode, to generate arbitrary
single-mode cavity field states \cite{eberly-engineering} and Fock states
with a large number of photons, through selective interactions \cite%
{marcelo-selective}. However, none of these schemes take into account the
system-environment interaction, which degrades the quantum states so that,
in general, the fidelity of the generated states decays quickly in time. To
circumvent this problem and to generate robust non-classical states of the
radiation field or of ionic motion, an approach based on \textit{reservoir
engineering} \cite{cirac1993, poyatos} has been proposed. This technique has
been employed in the trapped ion domain to protect, against decoherence and
relaxation, any superposition of Fock states \cite{carvalho-prl2001} in
one-dimensional motion of a single ion. Using similar procedures, there are
also proposals for the generation of trapping states (Fock states) of a
single cavity mode \cite{zelaquet1997} and entangled states of two cavity
modes, i.e., the two-mode squeezed vacuum state \cite{davidovich-eberly-li}.

In addition, schemes have been proposed to implement the interaction between
an atomic system and a squeezed reservoir, the simplest of which consists in
considering a two-level atom immersed in a squeezed multimode radiation
field \cite{scully-livro}. However, the scheme described does not represent
the interaction of an atomic system with an ideal squeezed reservoir, since
the action of the usual vacuum reservoir, due to the other modes of the
electromagnetic field, cannot be turned off and it is difficult to embed the
atom in a squeezed vacuum in a complete $4\pi $ solid angle. Parkins \textit{%
et al}. \cite{parkins1993} have shown how a two-level system can be coupled
to an almost ideal squeezed vacuum by assuming an atom strongly interacting
with a cavity field which is illuminated\textbf{\ }by finite-bandwidth
squeezed light. In Ref. \cite{lutkenhaus-pra1998}, the authors show how to
mimic the interaction of a two-level system with a squeezed reservoir
through quantum reservoir engineering. In this scheme, a four-level atom
interacts with circularly polarized fields. Then, assuming a strong decay of
the two most excited levels, it can be shown that the dynamics of the two
ground atomic states is effectively similar to that of a two-level system
interacting with a squeezed reservoir. We have also found a few experimental
studies of the dynamics of an atomic system in a squeezed reservoir \cite%
{kimble-squeezed-reservoir}, but these could not verify some predicted
phenomena, such as the phase sensitive decay of the atomic polarization,
mainly due to the difficulty of embedding atoms in a squeezed vacuum in a
complete $4\pi $ solid angle \cite{lutkenhaus-pra1998}.

In the present article we make a theoretical study of the manipulation of an
atom-field interaction and how to use engineered Hamiltonians to generate
both robust displaced squeezed states of a cavity field mode and a squeezed
reservoir for a two-level atomic system. To this end we employ the
interaction of three-level atoms in a $\Lambda $ configuration with a single
cavity mode and classical fields. Adjusting the intensity and the detuning
of the classical field, we derive an effective Hamiltonian which involves a
Jaynes-Cummings (JC), an Anti-JC and a rotation interaction of a two-level
atom with a cavity mode. With this kind of interaction, we can generate $i)$
a robust displaced squeezed state in a single cavity mode and $ii)$ an ideal
squeezed reservoir for atoms. In the next section, we present the model used
to obtain the desired Hamiltonian interaction. In section III, we show how
to use this interaction to generate a displaced squeezed state in a single
cavity mode and present a numerical analyzis of this system. In section IV,
assuming the atoms to be trapped in a bad cavity, we use the same effective
Hamiltonian to simulate an ideal squeezed reservoir for a two-level system.
We also carry out a numerical analyzis to verify the validity of
approximations employed to simulate a squeezed reservoir for atoms. Finally,
in section V we present some concluding remarks.

\section{The model}

To generate the desired effective interaction we will employ the interaction
of a three-level atom in a $\Lambda $ configuration with a single cavity
mode and classical fields. As depicted in Fig. 1, the ground $\left\vert
g\right\rangle $ and excited $\left\vert e\right\rangle $ states are coupled
to an auxiliary state $\left\vert i\right\rangle $ through classical fields,
with coupling $\Omega _{i}$ and frequency $\omega _{i}$ ($i=1-4$), and a
cavity mode, with coupling $g$ and frequency $\omega $. For this system, the
total Hamiltonian is $H=H_{0}+V\left( t\right) $, with 
\begin{subequations}
\label{eq1}
\begin{align}
H_{0}& =\hbar \omega _{g}\sigma _{gg}+\hbar \omega _{e}\sigma _{ee}+\hbar
\omega _{i}\sigma _{ii}+\hbar \omega a^{\dagger }a,  \label{eq1a} \\
V(t)=\hbar \left[ ga+\Omega _{1}\func{e}^{-i\omega _{1}t}+\Omega _{3}\func{e}%
^{-i\omega _{3}t}\right] \sigma _{ig}& +\hbar \left[ ga+\Omega _{2}\func{e}%
^{-i\omega _{2}t}+\Omega _{4}\func{e}^{-i\omega _{4}t}\right] \sigma
_{ie}+h.c.,  \label{eq1b}
\end{align}%
where $\hbar \omega _{\alpha }$, $\alpha =g,e,i$, are the energies of the
atomic levels, $\sigma _{lm}=\left\vert l\right\rangle \left\langle
m\right\vert $, $l,m=g,e,i$, are the atomic operators, $\ a$ and $a^{\dagger
}$ are the annihilation and creation operators for the cavity field,
respectively, and $h.c.$ stands for Hermitian conjugate. Using the unitary
transformation $U_{0}=$ $\func{e}^{-iH_{0}t/\hbar }$ we can re-write the
Hamiltonian in the interaction picture 
\end{subequations}
\begin{align}
H_{I}\left( t\right) & =\hbar \left[ ga\func{e}^{i\left( \Delta _{2}+\delta
_{2}\right) t}+\Omega _{1}\func{e}^{-i\left( \Delta _{1}+\delta _{1}\right)
t}+\Omega _{3}\func{e}^{i\left( \Delta _{3}+\delta _{3}\right) t}\right]
\sigma _{ig}  \notag \\
& +\hbar \left[ ga\func{e}^{-i\Delta _{1}t}+\Omega _{2}\func{e}^{i\Delta
_{2}t}+\Omega _{4}\func{e}^{i\Delta _{3}t}\right] \sigma _{ie}+h.c.,
\label{eq2}
\end{align}%
where we have defined $\Delta _{1}\equiv \omega -\left( \omega _{i}-\omega
_{e}\right) =\omega _{1}-\left( \omega _{i}-\omega _{g}\right) -\delta _{1}$%
, $\Delta _{2}\equiv \left( \omega _{i}-\omega _{g}\right) -\omega -\delta
_{2}=\left( \omega _{i}-\omega _{e}\right) -\omega _{2}$, and $\Delta
_{3}\equiv \left( \omega _{i}-\omega _{g}\right) -\omega _{3}-\delta
_{3}=\left( \omega _{i}-\omega _{e}\right) -\omega _{4}$. Considering the
non-resonant regime $\left\vert \Delta _{k}\right\vert \sim \left(
\left\vert \Delta _{k}\right\vert -\left\vert \Delta _{l}\right\vert \right)
\gg \left\vert g\right\vert \sqrt{\overline{n}},\left\vert \Omega
_{i}\right\vert $, $k\neq l=1,2,3$, $\overline{n}$ being the mean number of
photons in the cavity mode, we can adiabatically eliminate the transitions
between the ground/excited states and the auxiliary state, for example by
the methods described in \cite{James}. Thus, the effective dynamics,
considering only the atomic sub-space $\left\{ \left\vert g\right\rangle
,\left\vert e\right\rangle \right\} $, is governed by the effective
Hamiltonian 
\begin{align}
H_{eff}& =+\hbar \left\{ \left[ -\frac{\left\vert g\right\vert ^{2}}{\Delta
_{2}}a^{\dagger }a+\varpi _{g}\right] \sigma _{gg}+\left[ \frac{\left\vert
g\right\vert ^{2}}{\Delta _{1}}a^{\dagger }a+\varpi _{e}\right] \sigma
_{ee}\right\}  \notag \\
& +\hbar \left\{ \left[ \lambda _{1}a\func{e}^{i\delta _{1}t}+\lambda
_{2}a^{\dagger }\func{e}^{-i\delta _{2}t}+\beta \func{e}^{-i\delta _{3}t}%
\right] \sigma _{ge}+h.c.\right\} ,  \label{eq3}
\end{align}%
where $\varpi _{g}=\frac{\left\vert \Omega _{1}\right\vert ^{2}}{\Delta _{1}}%
-\frac{\left\vert \Omega _{3}\right\vert ^{2}}{\Delta _{3}}$, $\varpi _{e}=-%
\frac{\left\vert \Omega _{2}\right\vert ^{2}}{\Delta _{2}}-\frac{\left\vert
\Omega _{4}\right\vert ^{2}}{\Delta _{3}}$, $\lambda _{1}=\frac{g\Omega
_{1}^{\ast }}{\Delta _{1}}$, $\lambda _{2}=-\frac{g^{\ast }\Omega _{2}}{%
\Delta _{2}}$, $\beta =-\frac{\Omega _{3}^{\ast }\Omega _{4}}{\Delta _{3}}$.
For $\left\vert \Omega _{i}\right\vert \gg \left\vert g\right\vert $, the
dispersive atom-quantum field interactions are much smaller than the other
terms in the effective Hamiltonian. Therefore, under these conditions, we
make a new approximation, so that the effective Hamiltonian may be
re-written as 
\begin{equation}
H_{eff}\simeq +\hbar \left\{ \varpi _{g}\sigma _{gg}+\varpi _{e}\sigma
_{ee}\right\} +\hbar \left\{ \left[ \lambda _{1}a\func{e}^{i\delta
_{1}t}+\lambda _{2}a^{\dagger }\func{e}^{-i\delta _{2}t}+\beta \func{e}%
^{-i\delta _{3}t}\right] \sigma _{ge}+h.c.\right\} .  \label{eq4}
\end{equation}

By numerical analysis, we have verified the validity of this approximation.
We found that, the bigger the ratio $\left\vert \Omega /g\right\vert $, the
better were the results, as expected. Applying a new unitary transformation,
defined by the operator $U=\func{e}^{-i\left( \varpi _{g}\sigma _{gg}+\varpi
_{e}\sigma _{ee}\right) t}$, with the assumption $\delta _{1}=-\delta
_{2}=-\delta _{3}=\varpi _{e}-\varpi _{g}$, we can finally write the
effective Hamiltonian as 
\begin{equation}
H_{eff}\simeq \hbar \left\{ \left[ \lambda _{1}a+\lambda _{2}a^{\dagger
}+\beta \right] \sigma _{-}+h.c.\right\} ,  \label{Hefetivo}
\end{equation}%
with $\sigma _{-}=\sigma _{ge}$ and $\sigma _{+}=\left( \sigma _{-}\right)
^{\dagger }=\sigma _{eg}$. This effective Hamiltonian, which represents a
Jaynes-Cummings ($\lambda _{1}a\sigma _{+}+h.c.$) and an
anti-Jaynes-Cummings ($\lambda _{1}a^{\dagger }\sigma _{+}+h.c.$)
interaction, besides a rotation of the electronic states ($\beta \sigma
_{+}+h.c.$), can be used to carry out two distinct processes to generate $i)$
a robust displaced squeezed state for the radiation field and $ii)$ a
squeezed reservoir for an atom or an atomic sample. (A similar interaction
was employed in Ref. \cite{marcelo-selective} for the generation of large
Fock states through selective interactions.)

\section{Displaced squeezed state in a cavity mode.}

In this section, by a method similar to that used to generate a displaced
squeezed state in the trapped ion domain \cite{cirac1993}, we analyze the
generation of the same state for the radiation field trapped inside a high-$%
Q $ cavity, $\left\vert \alpha ,\varepsilon \right\rangle =D\left( \alpha
\right) S\left( \xi \right) \left\vert 0\right\rangle $, where $D\left(
\alpha \right) =\exp \left( \alpha a^{\dagger }-\alpha ^{\ast }a\right) $ is
the displacement operator, $\alpha $ being the amplitude of displacement,
and $S\left( \xi \right) =\exp \left[ \left( \xi ^{\ast }a^{2}-\xi
a^{\dagger ^{2}}\right) /2\right] $ is the squeezing operator, with $\xi
=re^{i\phi }$, $r$ and $\phi $ being the squeezing factor and squeezing
angle, respectively. To implement our proposal, an atomic beam should cross
the cavity under the action of classical fields in a way that the effective
interaction between each atom and the cavity mode is given by the effective
Hamiltonian (\ref{Hefetivo}). The atoms prepared in the ground state $%
\left\vert g\right\rangle ,$ are made to interact with the cavity mode
during a short time interval $\tau $ ($\lambda _{l}\tau \ll 1$, $l=1,2$), so
that the atomic beam acts as a reservoir at absolute zero ($T=0K$) for the
cavity mode, as described in various papers \cite{davidovich-eberly-li}.
Under these conditions, the steady state of the cavity field is exactly the
displaced squeezed state. To demonstrate this, we firstly apply a
time-independent unitary transformation to the effective Hamiltonian, as in
Ref. \cite{cirac1993}: $\widetilde{H}_{eff}=S\left( \xi \right) D(\alpha
)H_{eff}D^{\dagger }(\alpha )S^{\dagger }\left( \xi \right) $. In this
transformed representation, the Hamiltonian reads 
\begin{equation}
\widetilde{H}=\hbar \lambda a^{\dagger }\sigma _{-}+h.c.,
\label{Htransformado}
\end{equation}%
with the following adjustments: $\alpha \lambda _{1}+\alpha ^{\ast }\lambda
_{2}=-\beta $, $\tanh \left( r\right) =\left( \lambda _{1}/\lambda
_{2}\right) \func{e}^{-i\phi }$, and $\lambda \equiv \cosh \left( r\right)
\lambda _{2}-\func{e}^{-i\phi }\sinh \left( r\right) \lambda _{1}=\lambda
_{2}/\cosh \left( r\right) $. Here we can see that the squeezing factor $r$
is determined by the ratio $\left\vert \lambda _{1}/\lambda _{2}\right\vert $%
, and the amplitude of the coherent displacement, $\alpha $, by the
parameters $\beta $, $\lambda _{1}$, and $\lambda _{2}$. In this new
picture, the transformed Hamiltonian (\ref{Htransformado}) represents a
Jaynes-Cummings interaction between a cavity mode and a single two-level atom

As we can see from the diagram of levels in Fig. 1, the transitions $%
\left\vert g\right\rangle \longleftrightarrow \left\vert i\right\rangle $
and $\left\vert e\right\rangle \longleftrightarrow \left\vert i\right\rangle 
$ are dipole allowed and, by the selection rules, the transition $\left\vert
g\right\rangle \longleftrightarrow \left\vert e\right\rangle $ is not. In
this way, it would be very hard for the decay rate $\Gamma $ from the
excited state $\left\vert e\right\rangle $ to the ground state $\left\vert
g\right\rangle $ to be stronger than the effective atom-field coupling $%
\lambda $, so that we cannot use this channel of dissipation to engineer our
reservoir for the cavity mode, as in Ref. \cite{cirac1993}. To get round
this difficulty we can employ the scheme presented in Refs. \cite{morigi}
and \cite{davidovich-eberly-li} to simulate an atomic reservoir for the
cavity mode. To this end, we first assume that the atoms are initially
prepared in the ground state $\left\vert g\right\rangle $ and that the atoms
arrive in the cavity at the rate $r_{at}$. Next, we assume that each atom
interacts with the cavity field during a short time interval $\tau $, so
that $\lambda \tau \ll 1$. In this transformed picture the atom-field
interaction is governed by the transformed Hamiltonian (\ref{Htransformado}%
). Tracing on the atomic variables, the effective master equation for the
transformed cavity mode is given by \cite{morigi, davidovich-eberly-li} 
\begin{equation}
\frac{\partial \widetilde{\rho }}{\partial t}=\frac{\gamma _{eng}}{2}\left(
2a\widetilde{\rho }a^{\dagger }-a^{\dagger }a\widetilde{\rho }-\widetilde{%
\rho }a^{\dagger }a\right) ,  \label{eqmestra-campo}
\end{equation}%
where $\gamma _{eng}=r_{at}\lambda ^{2}\tau ^{2}$ is the engineered cavity
field decay rate. It is known that the vacuum state,$\left\vert
0\right\rangle $, is the steady state of Eq. (\ref{eqmestra-campo}) for the
cavity mode. Then, applying the reverse unitary transformation, we can
easily see that the steady state (for time $t\gg 1/\gamma _{eng}$) of this
system in the interaction picture is 
\begin{equation*}
\rho \left( t\rightarrow \infty \right) =D\left( \alpha \right) S(\xi )%
\tilde{\rho}S^{\dagger }(\xi )D^{\dagger }\left( \alpha \right) =D\left(
\alpha \right) S(\xi )\left\vert 0\right\rangle \left\langle 0\right\vert
S^{\dagger }(\xi )D^{\dagger }\left( \alpha \right) ,
\end{equation*}%
which is a pure state for the cavity mode, i.e., exactly the displaced
squeezed state $\left\vert \Psi \right\rangle =D\left( \alpha \right) S(\xi
)\left\vert 0\right\rangle $. The degree of squeezing $r$ is determined by
the amplitudes of the classical fields $\Omega _{j}$, since $\tanh \left(
r\right) =\left\vert \frac{\lambda _{1}}{\lambda _{2}}\right\vert $ and $%
\lambda _{1}=\frac{g\Omega _{1}^{\ast }}{\Delta _{1}}$ and $\lambda _{2}=-%
\frac{g^{\ast }\Omega _{2}}{\Delta _{2}}$. This steady state does not depend
on the initial cavity mode state: the generation of the displaced squeezed
state in the present scheme is achieved when the system reaches the steady
state. Here, the initial cavity field state only influences the time needed
for the system to achieve the steady state, as discussed in \cite%
{davidovich-eberly-li}. As pointed out in Ref. \cite{morigi}, the effective
master equation (\ref{eqmestra-campo}) for the cavity mode can be built even
for an atomic beam with random arrival times and without the need for atomic
detection nor precise interaction times between the atoms and the radiation
field. Hence, as in Ref. \cite{davidovich-eberly-li}, our scheme is robust
against stochastic fluctuations in the atomic beam and does not require
precise interaction times (velocity selection) or atomic detection.

To test the validity of the scheme we have perfomed a numerical solution of
the system. Starting with the cavity mode in the vacuum $\left\vert
0\right\rangle $ state and all the atoms in the ground state $\left\vert
g\right\rangle $, we carried out a numerical evolution of the system based
on Hamiltonian (\ref{eq3}). For simplicity, we fixed $\Omega _{3}=\Omega
_{4}=0$, which implies $\beta =0$ and a null displacement ($\alpha =0$). We
also chose the amplitudes $\Omega _{1}$ and $\Omega _{2}$ and the detunings $%
\Delta _{1}$ and $\Delta _{2}$ of the classical fields in such a way that $%
\lambda _{1}=0.1g$ and $\lambda _{2}\simeq 0.076g$, giving a squeezing
factor $r=1.0$, squeezing angle $\phi =0$, and $\lambda \simeq 0.065g$. The
interaction parameter was fixed at $\lambda \tau =0.2$, implying an
interaction time $\tau =0.2/\lambda \simeq 3.1/g$. With these adjustments,\
the evolution of the mean number of photons, $\left\langle n\right\rangle
=\left\langle a^{\dagger }a\right\rangle $, can be seen in Fig. 2.a, and
that of the variance of the cavity field quadratures $\left( \Delta
X_{l}\right) ^{2}=\left\langle X_{l}^{2}\right\rangle -\left\langle
X_{l}\right\rangle ^{2}$, $l=1,2$, $X_{1}=1/2\left( a+a^{\dagger }\right) $
and $X_{2}=-i/2\left( a-a^{\dagger }\right) $, in Fig. 2.b, both plotted
against the number of atoms that cross the cavity. In Fig. 2.a, \ for $r=1$,
the expected value for the mean number of photons of an ideal squeezed
state, $\left\langle n\right\rangle =\sinh ^{2}\left( r\right) =\sinh
^{2}\left( 1\right) \simeq \allowbreak 1.38$, is reached asymptotically. In
Fig. 2.b, the expected values for the variance in the quadratures of the
cavity field $\left( \Delta X_{1}\right) ^{2}=\exp \left( 2r\right) /4=\exp
\left( 2\right) /4\simeq \allowbreak 1.\,\allowbreak 85$ and $\left( \Delta
X_{2}\right) ^{2}=\exp \left( -2r\right) /4=\exp \left( -2\right) /4\simeq
0.034$ are also approached asymptotically. In Fig. 3 we have plotted the
Wigner function of the cavity field state: (a) for the initial state (vacuum
state) and then after the passage of (b) 50, (c) 100 and (d) 200 atoms
(steady state). Considering present-day technology \cite{haroche-RMP2001,
haroche-nature2007}, the cavity coupling strength $g\simeq 3\times 10^{5}$Hz
implies an interaction time per atom $\tau \simeq 3.1/g\simeq 10^{-5}s$ and
a total interaction time to reach the steady state around $200\times \tau
\approx 10^{-3}s$, which is almost three orders of magnitude smaller than
the current life-time of a photon inside a cavity ($\symbol{126}10^{-1}s$) 
\cite{haroche-nature2007, fock-state-cavity}.

\section{Squeezed vacuum reservoir for atoms.}

Our purpose in this section is to show how to simulate an ideal squeezed
vacuum for an atom or an atomic sample, trapped inside a bad cavity, whose
effective atom-cavity mode interaction is given by the same effective
Hamiltonian (\ref{Hefetivo}) employed in the last section for the generation
of a displaced squeezed cavity field state. In this case, the strong cavity
decay ($\Gamma $), compared to the other system parameters ($\lambda
_{1},\lambda _{2},\beta $), enables the atomic dynamics to be governed by an
effective Liouvillian identical to the squeezed vacuum reservoir for atoms.
Below we explain how this can be achieved. Firstly, turn off the classical
fields that generate rotations in the electronic states, i.e., $\Omega
_{3}=\Omega _{4}=\beta =0$, and re-write the effective Hamiltonian, Eq.(\ref%
{Hefetivo}), as%
\begin{equation}
H_{eff}\simeq \hbar \left( \lambda Ra^{\dagger }+\lambda ^{\ast }R^{\dagger
}a\right) ,  \label{Hefetivo2}
\end{equation}%
with $\lambda =\lambda _{2}/\cosh \left( r\right) $, as defined above, and $%
R=\cosh \left( r\right) \sigma _{-}-\sinh \left( r\right) \func{e}^{i\phi
}\sigma _{+}$. When the cavity decay is taken into account, the master
equation that governs the dynamics of the system, in the interaction
picture, is given by%
\begin{equation}
\overset{\cdot }{\rho }=-i\left[ H_{eff}\mathrm{,}\rho \right] +\frac{\Gamma 
}{2}\left( 2a\rho a^{\dagger }-a^{\dagger }a\rho -\rho a^{\dagger }a\right) +%
\mathcal{L}_{at}\rho {,}  \label{equacao_mestra-atomo}
\end{equation}%
where $\mathcal{L}_{at}\rho =\frac{\gamma }{2}\left( 2\sigma _{-}\rho \sigma
_{+}-\sigma _{+}\sigma _{-}\rho -\rho \sigma _{+}\sigma _{-}\right) $ stands
for the usual (weak) decay of a two-level system, $\gamma $ being the decay
rate of the atomic system, and $H_{eff}$ is given by Eq. (\ref{Hefetivo2})%
\textbf{.} To obtain the engineered reservoir, we take the decay constant of
the harmonic field to be significantly larger than the effective couplings, $%
\lambda _{1}$ and $\lambda _{2}$, and the decay constant $\gamma $ of the
two-level system. In our \textquotedblleft cavity QED +
atom\textquotedblright\ system, the regime $\Gamma \gg g$, $\gamma $ is
easily achieved with a cavity of low quality factor $Q$. Together with the
good approximation of a reservoir at absolute zero, the regime $\Gamma \gg g$%
, $\gamma $ enables us to consider only the matrix elements $\rho
_{mn}=\left\langle m\right\vert \rho \left\vert n\right\rangle $ inside the
subspace $\left\{ \left\vert 0\right\rangle \mathrm{,}\left\vert
1\right\rangle \right\} $ of Fock states. The equations of motion for the
elements $\rho _{mn}=\left\langle m\right\vert \rho \left\vert
n\right\rangle $ thus read 
\begin{align*}
\overset{.}{\rho }_{00}& =-i\left( \lambda ^{\ast }R^{\dagger }\rho
_{10}-\lambda \rho _{01}R\right) +\Gamma \rho _{11}+\mathcal{L}_{at}\rho
_{00}{,} \\
\overset{.}{\rho }_{10}& =-i\left( \lambda R\rho _{00}-\lambda \rho
_{11}R\right) -\Gamma /2\rho _{10}+\mathcal{L}_{at}\rho _{10}{,} \\
\overset{.}{\rho }_{11}& =-i\left( \lambda R\rho _{01}-\lambda ^{\ast }\rho
_{10}R^{\dagger }\right) -\Gamma \rho _{11}+\mathcal{L}_{at}\rho _{11}{,}
\end{align*}%
with $\overset{.}{\rho }_{01}=\left( \overset{.}{\rho }_{10}\right)
^{\dagger }$. Following the reasoning in Ref. \cite{carvalho-prl2001}, the
strong decay rate $\Gamma $ allows the adiabatic elimination of the elements 
$\rho _{01}$ and $\rho _{10}$. Tracing over the cavity field variables, the
atomic master equation reduces to

\begin{align}
\overset{\cdot }{\rho }_{at}& =\frac{\Gamma _{eng}}{2}\left( 2R\rho
_{at}S^{\dagger }-R^{\dagger }S\rho _{at}-\rho _{at}R^{\dagger }S\right) +%
\mathcal{L}_{at}\rho _{at}  \notag \\
& =\frac{\Gamma _{eng}}{2}\left\{ \left( N+1\right) \left( 2\sigma _{-}\rho
_{at}\sigma _{+}-\rho _{at}\sigma _{+}\sigma _{-}-\sigma _{+}\sigma _{-}\rho
_{at}\right) \right.  \notag \\
& +N\left( 2\sigma _{+}\rho _{at}\sigma _{-}-\rho _{at}\sigma _{-}\sigma
_{+}-\sigma _{-}\sigma _{+}\rho _{at}\right)  \notag \\
& \left. -2M\sigma _{+}\rho _{at}\sigma _{+}-2M^{\ast }\sigma _{-}\rho
_{at}\sigma _{-}\right\} +\mathcal{L}_{at}\rho _{at}\mathrm{,}
\label{eqmestra-atomo1}
\end{align}%
where $\Gamma _{eng}=4\left\vert \lambda \right\vert ^{2}/\Gamma $ stands
for the coupling strength of the engineered reservoir, $N=\sinh \left(
r\right) ^{2}$ and $M=\func{e}^{i\phi }\sinh \left( r\right) \cosh \left(
r\right) $. The inevitable and undesired action of the multimode vacuum $%
\mathcal{L}_{at}\rho _{at}$ thus works against the engineered reservoir for
the two-level system, leading to a non-ideal squeezed vacuum reservoir for
the atoms. However, in our proposal, as the transition $\left\vert
g\right\rangle \leftrightarrow \left\vert e\right\rangle $ is dipole
forbidden, these levels can be chosen in a way that the decay rate $\gamma $
can be very weak, so that $\Gamma _{eng}\gg \gamma $, making it possible to
neglect the term $\mathcal{L}_{at}\rho _{at}$ in Eq. (\ref{eqmestra-atomo1}%
). Therefore, with the present scheme we have achieved the interaction of an
ideal squeezed reservoir and a two-level atomic system. As pointed out in
Ref. \cite{lutkenhaus-pra1998}, this kind of interaction can produce some
interesting effects in atomic dynamics, such as suppression (enhancement) of
decay of the in-phase (out-of-phase) components of atomic polarization, and
line narrowing in resonance fluorescence and absorption spectra. Hence, the
present scheme could enable the observation of the effects predicted in the
context of squeezed bath--atom interactions, the properties of the squeezing
parameters, such as the (effective) photon-number expectation $N$ and the
squeezing phase $\phi $, being manipulated by the amplitude and phase of the
pumping fields that act on the atomic system. The equations of motion for
the expectation values of the operators $\sigma _{x}=\left( \sigma
_{-}+\sigma _{+}\right) $ and $\sigma _{y}=-i\left( \sigma _{-}-\sigma
_{+}\right) $ are%
\begin{align}
\left\langle \overset{\cdot }{\sigma }_{x}\right\rangle & =-\frac{\Gamma
_{eng}}{2}\left\{ \left[ 2N+2\left\vert M\right\vert \cos \left( \phi
\right) +1\right] \left\langle \sigma _{x}\right\rangle +2\left\vert
M\right\vert \sin \left( \phi \right) \left\langle \sigma _{y}\right\rangle
\right\} ,  \label{sigma-x} \\
\left\langle \overset{\cdot }{\sigma }_{y}\right\rangle & =-\frac{\Gamma
_{eng}}{2}\left\{ \left[ 2N-2\left\vert M\right\vert \cos \left( \phi
\right) +1\right] \left\langle \sigma _{y}\right\rangle +2\left\vert
M\right\vert \sin \left( \phi \right) \left\langle \sigma _{x}\right\rangle
\right\} .  \label{sigma-y}
\end{align}%
from which it can easily be shown that the atom has a phase-sensitive decay
when interacting with a squeezed vacuum reservoir \cite{scully-livro,
lutkenhaus-pra1998}. Therefore, the in-phase and out-of-phase components, $%
\left\langle \sigma _{x}\right\rangle $ and $\left\langle \sigma
_{y}\right\rangle $, of the atomic polarization decay at different rates,
depending on its initial phase relative to the phase $\phi $ of the
engineered reservoir. For an atom initially prepared in the eigenstate of
the operator $\sigma _{x}$, i.e. $\left\vert \Psi \right\rangle =1/\sqrt{2}%
\left( \left\vert g\right\rangle +\left\vert e\right\rangle \right) $, the
mean value $\left\langle \sigma _{x}\right\rangle $ evolves as%
\begin{equation}
\left\langle \sigma _{x}\left( t\right) \right\rangle =\frac{1}{2}\exp
\left( -\Gamma _{eng}\func{e}^{2r}t/2\right) \left[ 1+\cos \left( \phi
\right) \right] +\frac{1}{2}\exp \left( -\Gamma _{eng}\func{e}%
^{-2r}t/2\right) \left[ 1-\cos \left( \phi \right) \right] .
\label{sigma_medio}
\end{equation}%
We show the validity of our approximations leading to the dynamics of the
squeezed reservoir for atoms, by solving numerically Eq. (\ref%
{equacao_mestra-atomo}) with $H_{eff}$ given by Eq. (\ref{eq3}). In Fig. 4
we have plotted the evolution of $\left\langle \sigma _{x}(t)\right\rangle $%
, obtained from the approximate solution, i.e., Eq. (\ref{sigma_medio}),
which is based on an ideal squeezed reservoir for atoms, and its exact
(numerical) solution. We have fixed $r=1.5$, so that $\left\vert \lambda
_{1}/\lambda _{2}\right\vert =\tanh \left( 1.5\right) \simeq \allowbreak
0.90 $, \ \ $\lambda =\lambda _{2}/\cosh \left( 1.5\right) \simeq
\allowbreak 0.4\lambda _{2}$. In the exact solution we have also assumed $%
\left\vert \Omega _{1,2}\right\vert \sim 10\left\vert g\right\vert $ and $%
\left\vert \Delta _{1,2}\right\vert \sim 100g$ so that $\lambda \simeq
0.4g/10=0.04g\allowbreak $\ . We have assumed $\Gamma =40g$ and $\gamma =0$,
and found the evolution of $\left\langle \sigma _{x}\left( t\right)
\right\rangle $ for three different values of the squeezing angle, $\phi =0$%
, $\pi /2,$ and $\pi $. As we see in Fig. 4, the evolution of the $%
\left\langle \sigma _{x}\left( t\right) \right\rangle $ is phase dependent,
as expected for an ideal squeezed reservoir for atoms.

\section{Concluding remarks}

We have presented a theoretical study of the manipulation of the atom-field
interaction and its use in reservoir engineering. To build the desired
effective Hamiltonian we considered the interaction between an atomic system
in a $\Lambda $ configuration, driven by classical fields, and a single
cavity mode. With the engineered interaction, composed of interactions such
as Jaynes-Cummings, anti-Jaynes-Cummings and a rotation in the electronic
states, we firstly showed how to generate a decoherence-free displaced
squeezed state for the cavity field based on an atomic reservoir. In our
scheme an atomic beam works as a reservoir for the radiation field trapped
inside the cavity, as recently employed in Ref. \cite{davidovich-eberly-li}
to generate an Einstein-Podolsky-Rosen entangled radiation state in high-Q
resonators. Our scheme, as in Ref. \cite{davidovich-eberly-li}, is robust
against stochastic fluctuations in the atomic beam and does not require
precise interaction times (velocity selection) or atomic detection. Using
this system, we believe that a displaced squeezed cavity field state could
be experimentally generated with present-day technology. In addition, with
small changes, we were also able to generate an ideal squeezed reservoir for
two-level atomic systems \cite{scully-livro, lutkenhaus-pra1998}. For this
purpose, we had to assume, besides the engineered atom-field interaction, a
that the decay of the cavity field was much stronger than the effective
atom-field couplings. With this proposal some interesting effects in the
dynamics of an atom or an atomic sample in a squeezed reservoir can be
experimentally investigated. All the approximate theoretical results
presented in this work were checked by numerical analysis and all of them
showed excellent agreement with the exact (numerical) solutions.

\bigskip

\section{Acknowledgements}

We wish to acknowledge the support of the Brazilian agencies CNPq, CAPES,
FAPESP (process No. 2005/04105-5) and Brazilian Millennium Institute for
Quantum Information, and the help of T. J. Roberts in checking the
manuscript for grammatical errors. R. G. wishes to acknowledge support of
FONDECYT, (project No. 11060477).

\bigskip

\bigskip

\textbf{Figure Captions:}

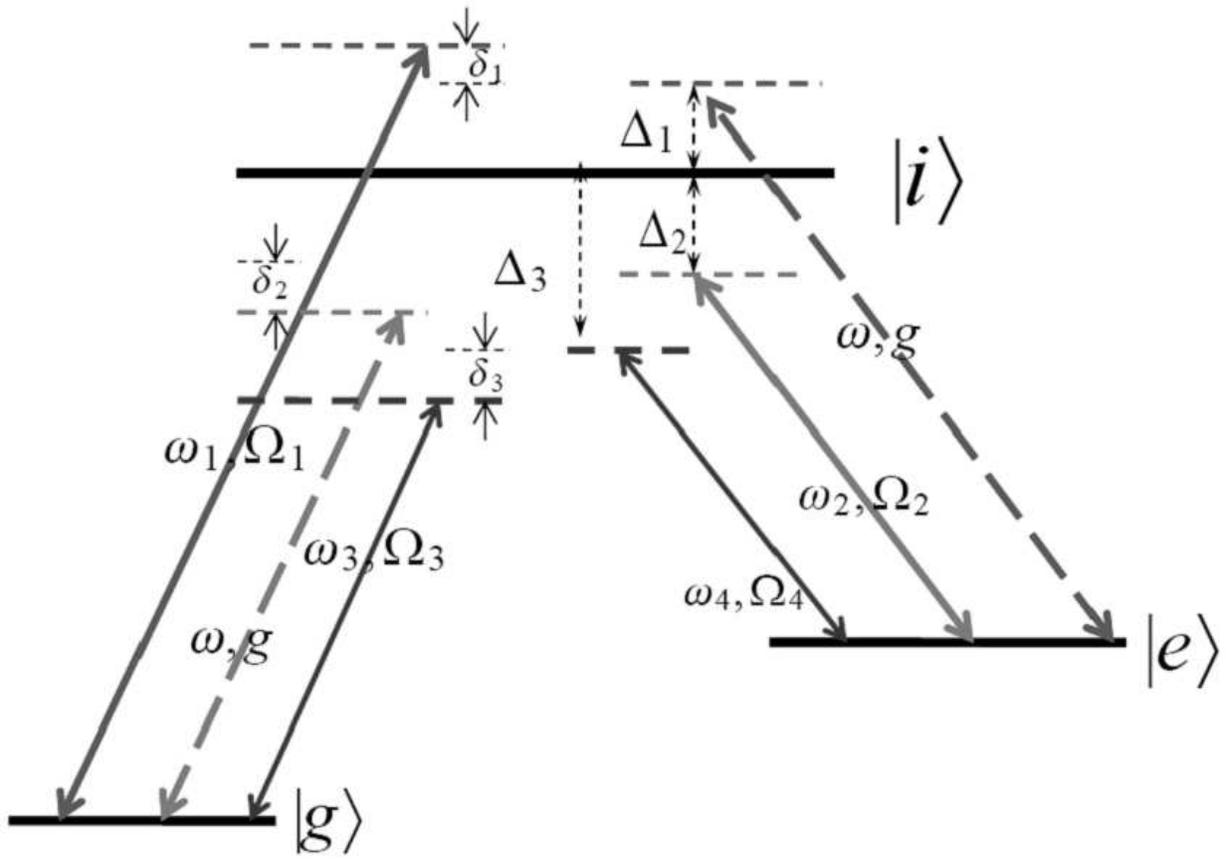
Fig. 1: Atomic and field configuration employed in the interaction
engineering process. The ground $\left\vert g\right\rangle $ and excited $%
\left\vert e\right\rangle $ states are coupled to the auxiliary state $%
\left\vert i\right\rangle $ through laser fields and the cavity mode.

Fig. 2: (a) Mean number of photons $\left\langle n\right\rangle
=\left\langle a^{\dagger }a\right\rangle $ and (b) Variance of the cavity
field quadratures, $\left( \Delta X_{l}\right) ^{2}=\left\langle
X_{l}^{2}\right\rangle -\left\langle X_{l}\right\rangle ^{2}$, $l=1,2$,
versus the number of atoms $N_{at}$ that cross the cavity, each of them
interacting during a time $\tau \simeq 3.1/g$. Squeezing factor $r=1.0$.
Solid line is obtained from the exact (numerical) solution of Eq. (\ref{eq3}%
) for a sequence of $N_{at}$ atoms. The dashed line represents the expected
(analytical) value.

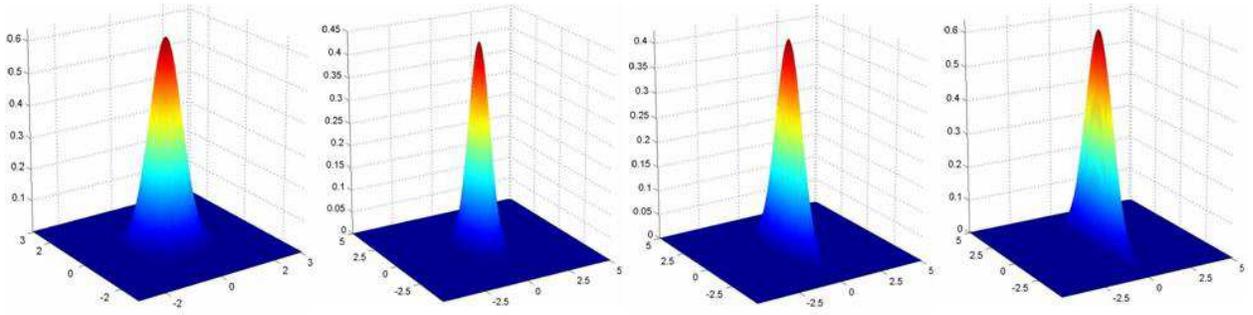
Fig. 3: Wigner function (and its projection) of the cavity field state: (a)
the initial (vacuum) state and after the passage of (b) 50, (c) 100 and (d)
200 atoms (steady state).

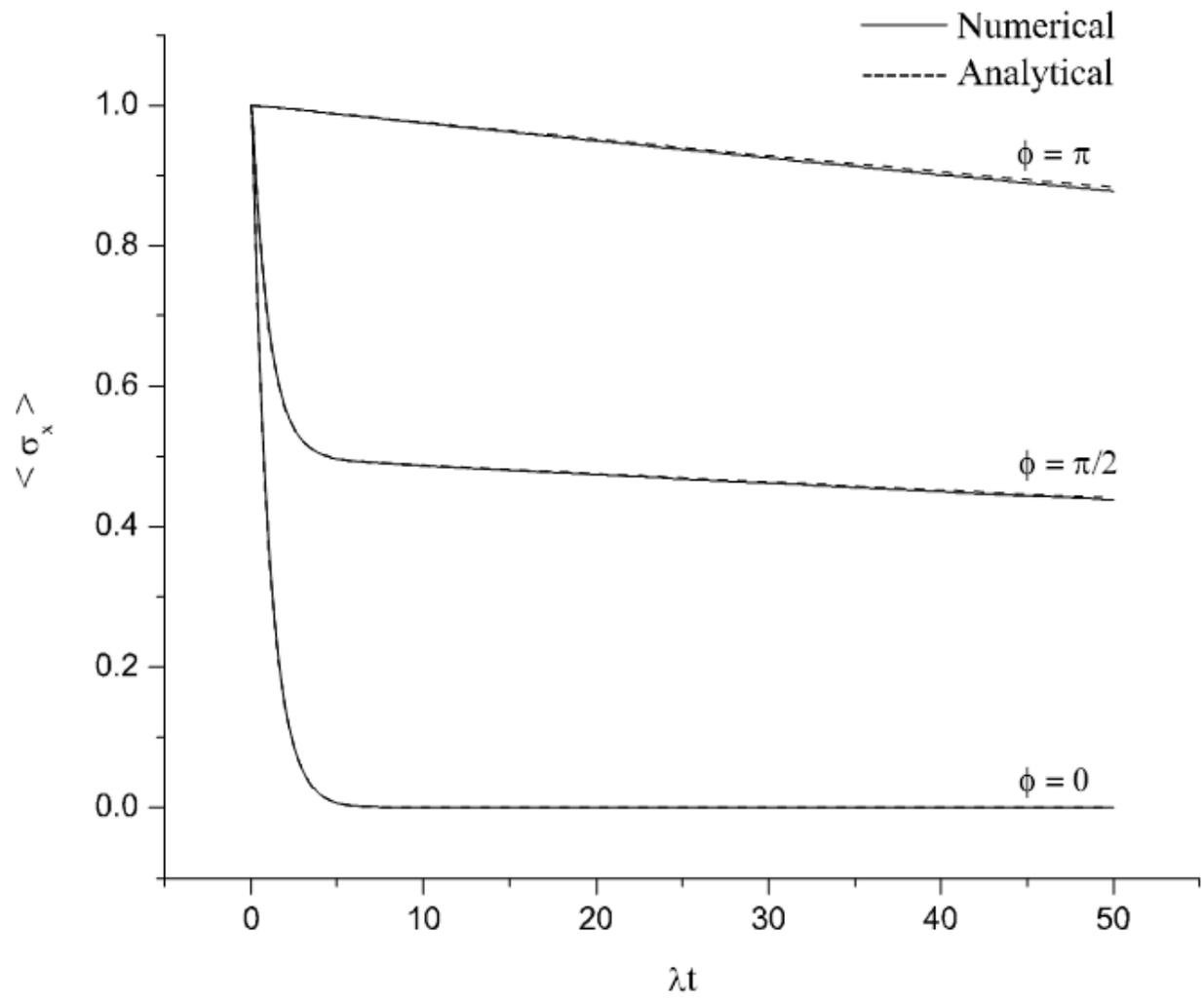
Fig. 4: Evolution of $\left\langle \sigma _{x}\left( t\right) \right\rangle $
for a squeezing factor $r=1.5$, $\Gamma =40g$, $\gamma =0$, and three
different values of the squeezing angle: $\phi =0$, $\pi /2,$ and $\pi $.
Solid line is obtained from the exact (numerical) solution of Eq. (\ref%
{equacao_mestra-atomo}) with $H_{eff}$ given by Eq. (\ref{eq3}). The dashed
line represents the expected (analytical) value, for an ideal squeezed
reservoir.

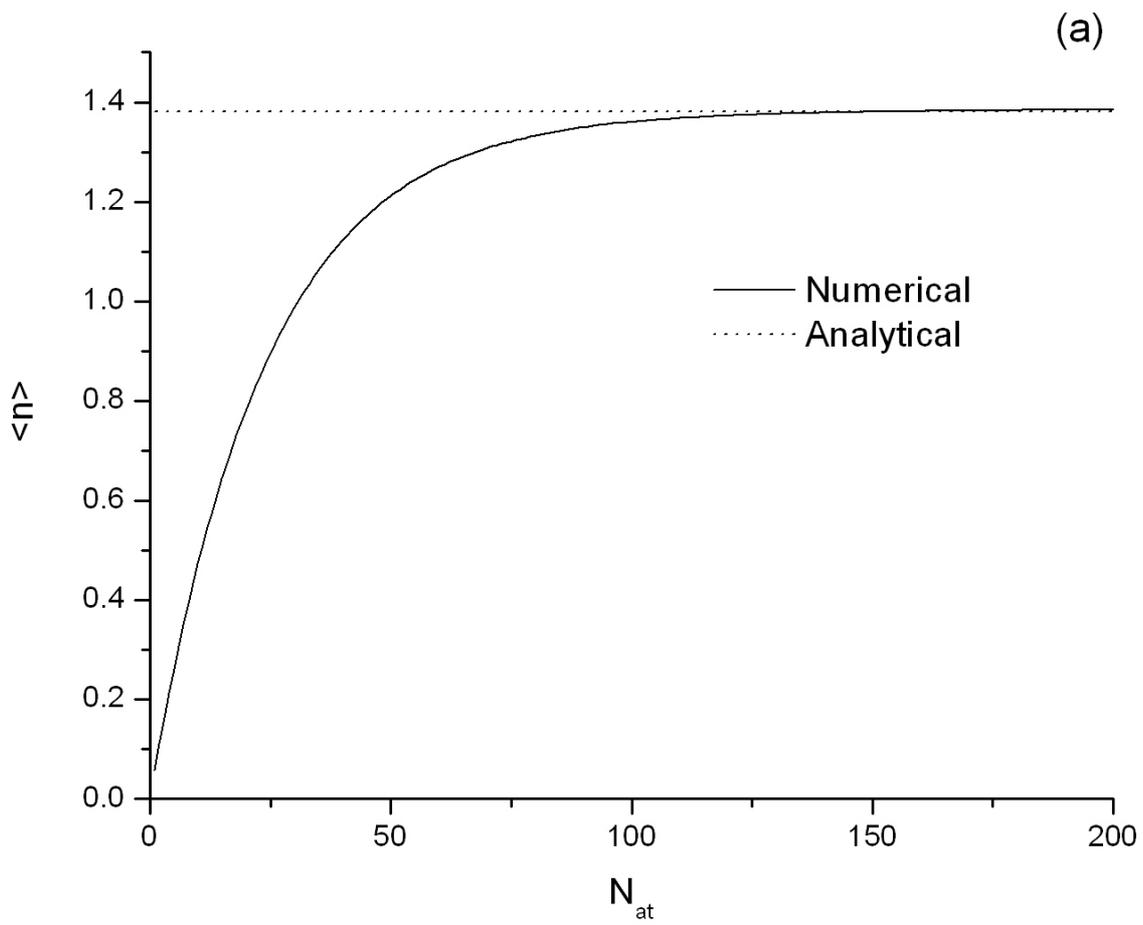
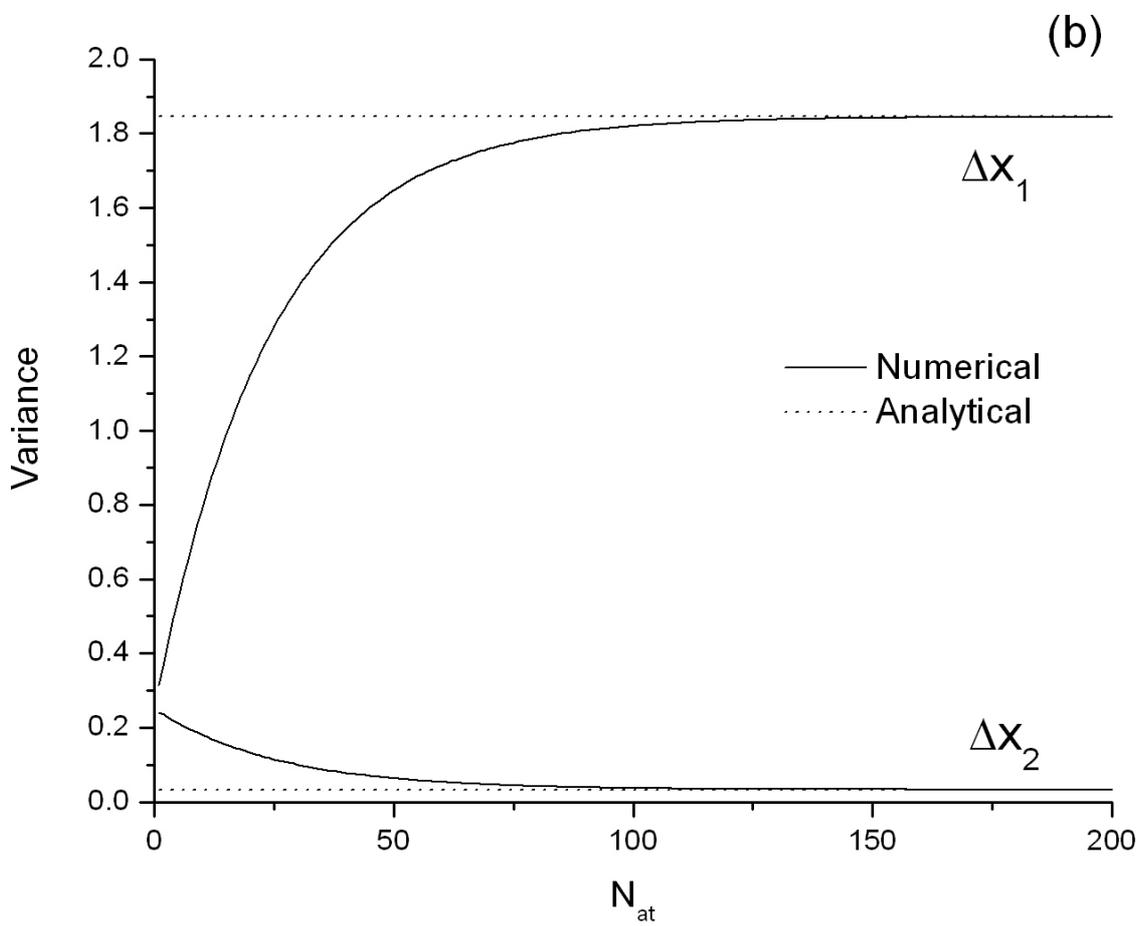
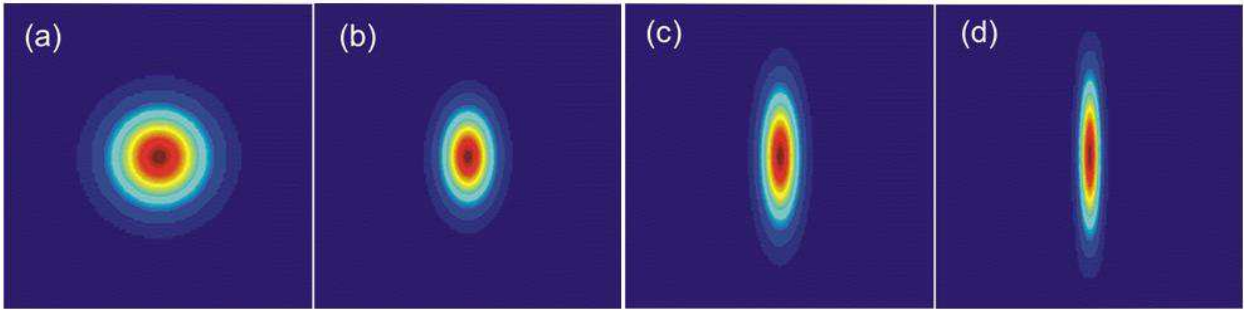
\end{document}